\documentclass[10pt]{amsart}
\usepackage{amsmath,amsfonts,amssymb,graphicx,color}
\usepackage[initials]{amsrefs}
\usepackage[all]{xy}
\usepackage{geometry}
\usepackage{latexsym}
\usepackage{multirow}
\usepackage{tikz}
\pgfsetlayers{background,main,foreground}
\usepackage{booktabs}
\usepackage{array}
\usetikzlibrary{calc,fit,shadows,arrows,positioning}
\pgfdeclarelayer{background}
\pgfdeclarelayer{foreground}
\pgfsetlayers{background,main,foreground}
\theoremstyle{plain}
\newtheorem{theorem}{Theorem}[section]
\newtheorem{remark}[theorem]{Remark}

\newtheorem{example}[theorem]{Example}

\numberwithin{equation}{section}

\newsavebox{\dataTableContent} 
\newenvironment{dataTable}[1] 
{%
	\begin{lrbox}{\dataTableContent}%
		\begin{tabular}{#1}}%
		{%
		\end{tabular}
	\end{lrbox}
	\begin{tikzpicture}
	\node [inner xsep=0pt] (tbl){\usebox{\dataTableContent}};
	\begin{pgfonlayer}{background}
	\draw[rounded corners=1pt,top color=gray!20,bottom color=gray!20,draw=black]
	(tbl.north east) rectangle (tbl.south west);
	\draw[rounded corners=1pt,top color=gray!10!black,bottom color=gray!50!black,draw=black]%
	($(tbl.north west)$) rectangle ($(tbl.north east)-(0,1.5\baselineskip)$);
	\draw[rounded corners=0.25pt,fill=gray,draw=black]%
	(tbl.south west) rectangle ($(tbl.south east)+(0,0.05)$);
	\end{pgfonlayer}
	\end{tikzpicture}}

\newenvironment{dataTable2}[1] 
{%
	\begin{lrbox}{\dataTableContent}%
		\begin{tabular}{#1}}%
		{%
		\end{tabular}
	\end{lrbox}
	\begin{tikzpicture}
	\node [inner xsep=0pt] (tbl){\usebox{\dataTableContent}};
	\begin{pgfonlayer}{background}
	\draw[rounded corners=1pt,top color=gray!20,bottom color=gray!20,draw=black]
	(tbl.north east) rectangle (tbl.south west);
	\draw[rounded corners=1pt,top color=gray!10!black,bottom color=gray!50!black,draw=black]%
	($(tbl.north west)$) rectangle ($(tbl.north east)-(0,2.5\baselineskip)$);
	\draw[rounded corners=0.25pt,fill=gray,draw=black]%
	(tbl.south west) rectangle ($(tbl.south east)+(0,0.05)$);
	\end{pgfonlayer}
	\end{tikzpicture}}

\tikzstyle{materia}=[draw, fill=gray!50, text width=6.0em, text centered,
minimum height=1.5em,drop shadow]
\tikzstyle{etape} = [materia, text width=4em, minimum width=0em,
minimum height=3em, rounded corners, drop shadow]
\tikzstyle{texto} = [above, text width=6em, text centered]

\tikzstyle{line} = [draw, thick, color=black!50, -latex']
\tikzstyle{ur}=[draw, text centered, minimum height=0.01em]

\begin{document}
\title[A simple unified explanation...]{A simple unified explanation of several genetic issues on today's human population and on archaic humans}
\author[Enflo]{Per H. Enflo}
\address[Per H. Enflo]{\newline \indent Department of Mathematical Sciences,\newline\indent  Kent State University, \newline\indent Kent, Ohio, 44242, USA.}
\email{per.enflo@gmail.com}
\author[Mu\~{n}oz]{Gustavo A. Mu\~{n}oz-Fern\'{a}ndez}
\address[Gustavo A. Mu\~{n}oz-Fern\'{a}ndez]{\newline \indent Instituto de Matem\'atica Interdisciplinar (IMI),\newline \indent Departamento de An\'{a}lisis Matem\'{a}tico y Matem\'atica Aplicada,\newline \indent Facultad de Ciencias Matem\'{a}ticas, \newline\indent  Universidad Complutense de Madrid,\newline\indent Plaza de Ciencias 3,\newline\indent  Madrid, 28040, Spain.}
\email{gustavo\_fernandez@mat.ucm.es}
\author[Seoane]{Juan B. Seoane-Sep\'{u}lveda$^{*}$}
\address[Juan B. Seoane-Sep\'{u}lveda]{\newline \indent Instituto de Matem\'atica Interdisciplinar (IMI),\newline \indent Departamento de An\'{a}lisis Matem\'{a}tico y Matem\'atica Aplicada,\newline \indent Facultad de Ciencias Matem\'{a}ticas, \newline\indent  Universidad Complutense de Madrid,\newline\indent Plaza de Ciencias 3,\newline\indent  Madrid, 28040, Spain.}
\email{jseoane@ucm.es}
\thanks{*Corresponding author.}
\begin{abstract}
We will give a simple, unified, possible explanation of several debated genetic issues on today’s humans, Neandertals and Denisovans. In particular it is shown by means of a simple mathematical model why there is little genetic variation in todays's human population or in Western Neandertal population, why all mtDNA and y-chromosomes in today's humans seem to have African origin with no trace of Neandertal nor Denosovan mtDNA or y-chromosomes, why a big part of the European gene pool is young (from Neolitic time), and why today's East Asians have mode Neandertal genes than today's Europeans.
\end{abstract}
\maketitle
\tableofcontents
\section{Introduction}

Observations (1.)-(7.) below are debated today and several different explanations have been suggested. In this paper we propose a simple unified explanation for all these observations.

\begin{itemize}
	{\color{black}	\item[(1.)]  The genetic variation in today's human population is surprisingly small. Most of the genetic material in today's human population has African origin. This has sometimes been attributed to a population bottleneck some 50,000-100,000 years ago.}
	\item[(2.)] There was small genetic variation in the late western Neandertal population. This has sometimes been attributed to a population bottleneck. 
	\item[(3.)] All mtDNA in today's human population seems to have African origin. There seems to be no Neandertal or Denisovan mtDNA in today's human population.
	\item[(4.)] All y-chromosomes in today's human population seem to have African origin. There seems to be no Neandertal or Denisovan y-chromosome in today's human population.	
	\item[(5.)] A big part of the European gene pool is young - from Neolithic time.
	\item[(6.)] Today's East Asians have more Neandertal genes than today's Europeans.
	{\color{black}
	\item[(7.)] It is easier to find neandertal genes in today's human population, than to find genes of moderns in the neandertal population.}

\end{itemize}

We will see below that, in order to explain (1.)-(7.) above, it is enough to just assume that there have been``reproductively advantageous regions'' (resp. ``reproductively disadvantageous regions'') in the world and that this has led to migration between regions. We will see here that there is evidence to support that assumption.

We do not need to assume any superiority or inferiority in any way of any group of people. And we can just assume total and random interbreeding between different kinds of people (Moderns, Neandertals, Denisovans) and that couples, on average, give birth to $50\%$ males and $50\%$ females.  We do not need to assume lower fertility in the admixture of different kinds of people. Lower amount of interbreeding or lower fertility in interbreeding between different kinds of people, will sometimes strengthen, but mostly not affect our conclusions. We will discuss such situations. Our explanation of (1.) and (2.) is consistent with a population bottleneck but will not require any population bottleneck.

The paper is organized as follows. In Section 2 we discuss how genetic material disappears in a reproductively disadvantageous region. It is exponentially fast, i.e., much faster than the disappearance by random genetic drift. In Section 3 we discuss how unstable and fluctuating living conditions make a region reproductively disadvantageous. We give numerical estimates.  And, as we will see, from just Sections 2 and 3 we draw the following conclusion:

\textit{If we just assume that in the last 50,000-100,000 years, Europe has had the least stable living conditions, Asia has had more stable living conditions and Africa has had  the most stable living conditions, then we should expect to make the observations (3.)-(7.) above.}

In Section 4 we will discuss how the genetic material from a reproductively advantageous region will, over time, be dominating in a much larger region. We will give numerical estimates on how fast this happens. We will see, that a scenario with a population bottleneck is just a special case of the more general scenario, where we have a small reproductively advantageous region. And we will see that over time, these scenarios will end up looking genetically similar. So, from Section 4 we will get the following conclusion:

\textit{If we assume that in the last 50,000-100,000 years, there has been a region in Africa with more stable living conditions than in other parts of the world, then we should expect  that most of the genetic material in today’s human population has its origin in that region, i.e., we should expect  observation (1.) above.}

\textit{Our explanation of observation (2) is analogous. It is enough to assume that some region in western Europe was more reproductively advantageous than the others.}

\begin{remark}\label{rem01}
{\rm
The concept of ``reproductively disadvantageous regions'' was originally introduced (1999) in an attempt to reconcile the evidence of Neandertal ancestry in today’s human population (presented, e.g., in Duarte et al. \cite{Trinkausetal}, Zilh\~ao and Trinkaus \cite{Trinkaus} and Frayer \cite{Frayer}) with the genetic information (Krings et al. \cite{NEW-5}) indicating that there is no Neandertal mtDNA in today’s human population and that, at that time, there was little evidence of any Neandertal DNA in today's human population. In Enflo et al. \cite{Enflo}, a short version of the model is presented.
}
\end{remark}

\begin{remark}\label{rem02} 
{\rm Obviously, the fact that observations (1.)-(7.) can be explained from some simple assumptions does not in any way contradict that several other phenomena are essential parts of a complete understanding of these observations.}
\end{remark}

\section{The disappearance of genes in in a reproductively disadvantageous region} \label{sec:3}

We say that a region is reproductively disadvantageous if, on average, every couple has $(1-\epsilon)$ female and $(1-\epsilon)$ male fertile offspring, with $0 < \epsilon < 1$, and the population size is kept up by immigration. In such a region, after one generation, $(1-\epsilon)$ of the original mtDNA carried by females, remains in the region, and  $(1-\epsilon)$ of the original y-chromosomes carried by males remain in the region. Assuming that the fertility of the immigrants and the fertility of the admixture between immigrants and original local population are the same as the fertility of the original local population - then, after two generations,  $(1-\epsilon)^2$ of the original mtDNA carried by females remains in the region and $(1-\epsilon)^2$ of the original y-chromosomes carried by males remain in the region. If the fertility of the admixture between immigrants and local population is lower, then even less original mtDNA and y-chromosomes will remain.
After $n$ generations, 
\begin{equation}\label{eq:estimate1}
(1-\epsilon)^n
\end{equation}
of the original mtDNA carried by females and $(1-\epsilon)^n$ of the original y-chromosomes carried by males will remain in the region. 

Estimate \eqref{eq:estimate1} is valid regardless of the amount of interbreeding between immigrants and local population. In particular, it is valid whether there is total and random interbreeding or no interbreeding between local population and immigrants. Obviously, in the latter case, in every generation, an individual will either carry only genes from the original local population or carry only genes from the immigrants. Estimate \eqref{eq:estimate1} also holds for genes, neutral to natural selection.

\begin{remark}\label{rem21}
{\rm The disappearance of mtDNA, y-chromosomes and genes neutral to natural selection, is (mathematically) analogous to the following situation: Consider a bottle, filled with a fluid different from water (corresponds to the original local population). Assume that the bottle is leaking an $\epsilon$-proportion of its content per unit of time and is replenished with water (corresponds to the immigrants) to keep it full. Assume that the proportions of original fluid and water, that is leaking out, are the same as in the bottle. Then, after n time units, $(1-\epsilon)^n$ of the original fluid remains in the bottle. And this estimate is independent of how much or how little the fluid and the water mix with each other.}
\end{remark}

We now give a numerical example to show the effect of the estimate \eqref{eq:estimate1}. We will assume that one generation is 23 years.

\begin{example}\label{ex13}
{\rm Consider a region that is initially populated by $100,000$ males and $100,000$ females. Assume that in the region, on average, every male and every female will have $0.98$ male offspring and $0.98$ female offspring. Assume that the population size is kept up by immigration. Then, by estimate \eqref{eq:estimate1}, after n generations, $0.98^n$ of the original mtDNA and $0.98^n$ of the original y-chromosomes remain in the region. That is, after $35$ generations (approximately $800$ years) half of the original mtDNA and half of the original y-chromosomes remain. And after 350 generations (approximately 8,000 years) less than 0.001\% of the original mtDNA and the original y-chromosomes remain. After $15,000$ years, with more than $99\%$ probability, one cannot find a single individual with mtDNA or y-chromosome from the original population. This holds, independently of how much or little interbreeding there is between immigrants and local population.}
\end{example}

\begin{remark}\label{rem14}
{\rm We may think of Example \ref{ex13} as mimicking a situation where the initial people in the reproductively disadvantageous region are Neandertals or Denisovans and the immigrants are Modern Humans. Now, relatively recent findings (see references \cite{Paabo2014} and \cite{Sankararaman2016}) suggest that there may have been lower fertility in male admixtures between, on the one hand Neandertals or Denisovans, and on the other hand Modern Humans. References \cite{Paabo2014} and \cite{Sankararaman2016} suggest that this can be caused by male hybrid sterility genes in the X-chromosome. Since male hybrid sterility genes will only survive in female offsprings, such genes in a local population will disappear fast in a region, that is constantly replenished with X-chromosomes from immigrants, which do not carry these male hybrid sterility genes. So, for this situation, the disappearance of  mtDNA or y-chromosomes or selectively neutral genes of a local population will be at least as fast as the estimate \eqref{eq:estimate1}.
}
\end{remark}

\begin{remark}\label{rem15}
{\rm In a population, not all genetic material will be passed on from one generation to the next. This disappearance of genes, known as “random genetic drift” is, however, much slower than the disappearance given by \eqref{eq:estimate1}  above. The disappearance given by \eqref{eq:estimate1} is exponential in time, whereas the disappearance by random genetic drift is inversely linear in time. For more details about random genetic drift we refer the reader to \cite{Hartl} (chapters 5 and 7).}
\end{remark}

\begin{remark}\label{rem17} 
{\rm If a small population in a region mixes with a much larger population in the region then a randomly chosen gene from the region will very likely come from the large population. This holds, independently of the amount of interbreeding between the populations. In a small sample of genes, it is likely that they will all come from the larger population, but it is unlikely in a large sample. For instance, if the larger population is 100 times as big as the small population then, for a sample of 100 genes, the probability that they will all come from the large population is $$(100/101)^{100} \approx 1/e \approx 0.37.$$ However, in a sample of 1,000 genes, the probability that they will all come from the large population is less than $0.0001$.}
\end{remark}
	
\begin{remark}\label{rem18} 
{\rm Genes for which there is a selective advantage may survive in a reproductively disadvantageous region. Obviously, this can only happen if there is some interbreeding between immigrants and local population. The bigger the reproductive disadvantage of the region is, the fewer genes can be expected to survive.}
\end{remark} 

\section{Examples of reproductively disadvantageous regions}\label{sec:4}
Regions with unstable, fluctuating living conditions are often examples of  reproductively disadvantageous regions. This is due to the fact that, in bad times, the population size will go down (mostly by deaths and lower reproductivity). And, in good times, the population size will go up again, not just by increased reproductivity but also by immigration. Examples \ref{ex21}, \ref{ex22}, \ref{ex23} and \ref{ex24} are such examples.

\begin{example}\label{ex21}
{\rm Cities, both modern and ancient, are examples of reproductively disadvantageous regions. Cities are usually good places to live, and people move into them. However, sometimes, there are bad events (such as epidemics or fires) which make the population size go down. After the bad event, the population size is restored, not just by increased reproductivity, but also by immigration.}
\end{example}

\begin{example}\label{ex22}
{\rm Assume that in a region, every five generations (115 years) there is a bad event (like a very long winter) which kills $20\%$ of the population. Assume that after the bad event the population is restored, $10\%$ by increased reproduction and $10\%$ by immigration. Then, the rate  of disappearance of mtDNA, y-chromosomes or genes neutral to natural selection will be the same as in Example \ref{ex13} above.}
\end{example}

\begin{example}\label{ex23}
{\rm Assume that (because of uneven supply of food) the carrying capacity of a region has a maximum of 10,000 people and a minimum of 5,000 people in every period of 150 years. Assume that there is no emigration out of the region. Assume that every time the population increases from 5,000 to 10,000, 2,000 is due to increased reproductivity and 3,000 is due to immigration. If the region has initially 10,000 people then, after 150 years, only $70\%$ of the original mtDNA, y-chromosomes or selectively neutral genes remain in the region and after 300 years less than $50\%$ remain in the region. After 3,000 years, less than $0.1\%$ of the original mtDNA, y-chromosomes or selectively neutral genes remain in the region, and after 6,000 years, with more than $99\%$ probability, one cannot find one single individual with mtDNA or y-chromosome from the original population.}
\end{example}

\begin{example}\label{ex24} 
{\rm There is evidence that Europe in the last 50,000-100,000 years has been a region with  unstable, fluctuating living conditions and that population size after the bad periods has partly been restored by immigration. Following the work in \cite{Morin}, there was a population contraction in the period 40,000-35,000 years BP, and something similar around the glacial maximum, in the period around 18,000 years BP, long after the Neandertals had disappeared. Besides these big fluctuations, there has likely been many smaller fluctuations, even in the last 15,000 years. There is evidence that living conditions in Asia have been more stable than in Europe and in Africa even more stable than in Asia. From this it follows that, relative to Africa, Asia has been reproductively disadvantageous and Europe even more reproductively disadvantageous. The computations above indicate that even a small reproductive disadvantage will, over time, lead to the disappearance of mtDNA, y-chromosomes and selectively neutral genes.}
\end{example}

\textit{Example \ref{ex24} explains why all mtDNA and all y-chromosomes in today’s human population seem to have African origin.  And it explains why most of today’s European gene pool is of  Neolithic origin (Chikhi et al., \cite{NEW-11}).
We see that the more stable living conditions in Asia than in Europe has made Asia less reproductively disadvantageous than Europe and, by Remark \ref{rem18} above, that explains why more Neandertal genes have survived in today’s East Asians than in today’s Europeans. (Wall et al., \cite{NEW-10}).  Thus, Example \ref{ex24} gives a possible explanation of the observations (3.)-(6.) above.}

\begin{example}\label{ex25}   
{\rm In many regions of the world today, the populations do not reproduce themselves but depend on immigration to keep up the population size. There is a big area where, on average, every individual has no more than $1.6$ children.}
\end{example}
{\color{black}
Regarding observation (7.), the small amount of neandertal genes in today's human population has been attributed to a relatively small amount of interbreeding events between moderns and neandertals. Our model shows that we get the same genetic outcome also under a scenario of total and random interbreeding which goes on continuously in time. It is only the neandertal genes for which there is positive selection, that then survive, since the region(s) where the neandertal lived were reproductively disadvantageous.
This last scenario provides a possible explanation, why we do not as easily see an introgression of genes of moderns into the neandertal population. In the last 60000 years there was no such introgression, caused by a limited number of mating events.The flow of genes from moderns into the neandertal population, in the last 60,000 years, dissolved the neandertal population. It changed the neandertals to being moderns (see, also, \cite{nueva}). }

\section{Model of genetic flow from a small, reproductively advantageous region, into a much larger region. Applications to issues on bottlenecks in human populations and to issues on Neandertals}

In this section we will only consider genes neutral to natural selection. In the previous sections we have considered what happens in a region that is reproductively disadvantageous, and where there is immigration into the region to keep up the population size. In this section we will study a different and somewhat more complicated situation. We will have a region with a reproductive surplus that leads to migration. We will see that even if that region is fairly small and the reproductive surplus is fairly modest, over time, genes from that region will dominate much larger areas. 

Consider $n$ regions $R(1),R(2),\ldots,R(n)$, with initial population sizes ($=$ amount of genes) $p(1)$, $p(2)$, $\ldots, p(n)$. We assume that $R(1)$ has a reproductive surplus. On average, every individual in $R(1)$ has $1+r$ male offspring and $1+r$ female offspring, $0<r<1$. We assume that $R(2),R(3),R(3),\ldots,R(n-1)$ are reproductively neutral, that is, in each of these regions, on average, every individual has one male offspring and one female offspring. We assume that $R(n)$ is reproductively disadvantageous, that is, in $R(n)$, on average, every
individual has $1-r$ male offspring and $1-r$ female offspring. We assume that in every generation, the proportion $r$ of the population in
$R(m)$ migrates to $R(m+1)$. We assume that in every generation and that the probability for a gene to reproduce and the probability for a gene to travel from $R(m)$ to $R(m+1)$ is independent of where the gene originally came from.

\begin{remark}\label{rem31}
{\rm The assumptions made in the previous paragraph can be fulfilled, whether or not, in any generation, there is no mating, some mating and reproduction, random mating and reproduction between individuals who have migrated from $R(m)$ to $R(m + 1)$ and those who were already in $R(m+1)$. The assumptions will not be fulfilled if admixtures between individuals from different regions have lower fertility. Considerations like those above suggest that the dominance of genes from R(1) will be at least as large in that case.} 
\end{remark}

Under the assumptions considered just before Remark \ref{rem31}, the change in genetic material from one generation to the next is given by the $n \times n$ transition matrix $M(r)$ given by

$$
M(r)=\left(
\begin{array}{cccccccc}
1     &    0   &  0  &  0  & \ldots  &    0   &  0  &  0\\
r & 1-r & 0  &  0 & \ldots  &    0   &  0  &  0\\
0 & r & 1-r  &  0 & \ldots  &    0   &  0  &  0\\
0 & 0 & r  &  1-r & \ldots  &    0   &  0  &  0\\
\vdots & \vdots & \vdots  &  \vdots & \ddots  &    \vdots   &  \vdots  &  \vdots\\
0 & 0 & 0  &  0 & \ldots  &    1-r   &  0  &  0\\
0 & 0 & 0  &  0 & \ldots  &    r   &  1-r  &  0\\
0 & 0 & 0  &  0 & \ldots  &    0   &  r  &  1-r\\
\end{array}
\right).
$$

The entry $ij$ in $M(r)$ represents the proportion of genes in $R(j)$ which migrate to $R(i)$ in one generation. The entry $jj$ is the proportion of genes in $R(j)$ which stay in $R(j)$. The entry $ij$ in the matrix $M(r)^k$ is the proportion of genes from $R(j)$ which travel to $R(i)$ in $k$ generations.

The column $m$, $1 < m < n$, in $M(r)$ adds up to $1$, which represents the reproductive neutrality of $R(m)$. The first column in $M(r)$ adds up to $1+r$ and the last column to $1-r$, which represents the reproductive advantage of $R(1)$ and the reproductive disadvantage of $R(n)$.

To study the amount of genes in each region after 1 generation we multiply $M(r)$ by the column 
$$
P(0)=\left(
\begin{array}{c}
p(1)\\
p(2)\\
p(3)\\
\vdots\\
p(n)\\
\end{array}
\right),
$$
which gives us the column
$$
\left(
\begin{array}{c}
p(1)\\
r p(1) + (1-r) p(2)\\
rp(2)+(1-r)p(3)\\
\vdots\\
rp(n-1) + (1-r) p(n)\\
\end{array}
\right).
$$

Thus, after $1$ generation, $R(1)$ has $p(1)$ genes which all originated in $R(1)$, $R(2)$ has $r p(1) + (1-r) p(2)$ genes, of which $r p(1)$ originated in $R(1)$ and $(1-r) p(2)$ 
originated in $R(2)$, and so on.
It is easy to verify  (both theoretically and numerically) that if we repeat this process, that 
is, multiply the matrix $M(r)^k$ by the column $P(0)$ we will get, in the limit as $k$ 
tends to infinity, the column 
$$
P=\left(
\begin{array}{c}
p(1)\\
p(1)\\
p(1)\\
\vdots\\
p(1)\\
\end{array}
\right),
$$
that is, in the limit, every region has the 
amount $p(1$) of genes, and all genes come from $R(1)$.      

Thus, by the previous comment, no matter what the initial population sizes are in the regions $R(2), R(3), \ldots, R(n)$, the final result given above is the same.

It is also easy to verify that, after $k$ generations, the amount of genes in $R(m)$, $1\le m \le n$, which have originated in $R(1)$, depends on $p(1)$ but not on $p(2)$, $p(3),\ldots, p(n)$. Thus, for the amount of genes in $R(m)$ which have originated in $R(1)$, the bottleneck case is similar to the case without a bottleneck.

We will now study 2 scenarios, in order to see how fast is the convergence to the limit situation. Scenario 1 is the bottleneck scenario, where (initially) only $R(1)$ is populated. So, $p(1) \ne 0$, $p(2)=p(3)= \ldots =p(n)=0$. In Scenario 2, all regions are initially populated with the same population size (same amount of genes) in all regions. So, $p(1)=p(2)= \ldots =p(n)$.

In Scenario 1, the total amount of genes in all regions will grow to the limit case , when there are $p(1)$ genes in each region, all originating in $R(1)$. In Scenario 2, the amount of genes in each region will remain constant $=p(1)$ over the generations. The amount of genes originating in $R(1)$, in the region $R(m)$, $1\le m \le n$, is, for all $k$, after $k$ generations, the same in Scenario 1 and Scenario 2.
 
In Table 1, we study Scenario 1 and Scenario 2. We have 200 regions, i.e., $n=200$, and the reproductive surplus in $R(1)$ has $r = 0.1$, i.e., on average, every individual in $R(1)$ will have $1.1$ male and $1.1$ female offspring, and, in every generation $0.1 p(1)$ genes will migrate from $R(1)$ to $R(2)$. And, more generally, in Scenario 2, for $1<m<n$, in every generation, the amount of $0.1 p(1)$ genes will migrate from $R(m)$ to $R(m+1)$, but all of them have not originated in $R(1)$. $R(n)$ will have a reproductive deficit of $0.1$ in each generation. So, in Scenario 2, the amount of genes in each region will be $p(1)$ all the time. Table 1 shows that after 200 generations ($=200 \cdot 23 = 4,600$  years) genes originating in $R(1)$ will make up more than $99\%$ of the genes in the regions $R(1), R(2), \ldots, R(12)$. And Table 1 shows, that after $2,400$ generations ($=2,400 \cdot 23 = 55,200$ years), genes originating in $R(1)$ will make up more than $99\%$ of the genes in each of the regions $R(1), R(2),\ldots, R(200)$.

Table 1 also gives information about Scenario 1, the bottleneck scenario. If we start with $p(1)$ genes in $R(1)$, and $0$ genes in $R(2), R(3), \ldots, R(200)$, and as above, $r=0.1$, $n=200$, the second column of Table 1 shows, that after $200$ generations ($=200 \cdot 23=4,600$ years), we have more than $0.99 p(1)$ genes in each of the regions $R(1), R(2), \ldots , R(12)$, all genes originating in R(1). And the second column of Table 1 shows, that after $2,400$ generations ($=2,400\cdot 23=55,200$ years), we have more than $0.99 p(1)$ genes in each of the regions $R(1), R(2),\ldots , R(200)$, all genes originating in $R(1)$. So Scenario 1 and Scenario 2 will end up looking very similar.
{\small\begin{table}[h!]
	\begin{center}
		\begin{dataTable2}%
			{@{\hspace{2ex}} c @{\hspace{2ex}} c @{\hspace{2ex}} c @{\hspace{2ex}} c @{\hspace{2ex}} c @{\hspace{2ex}} c@{\hspace{2ex}} c@{\hspace{2ex}} c @{\hspace{2ex}}}%
			
			\multirow{2}{2.5cm}{\textcolor{white}{After this many\\ generations}} &
			 \multicolumn{6}{c}{\textcolor{white}{The number of regions where the amount of genes originating in $R(1)$ is}}\\
			\cline{2-7}
			& \textcolor{white}{$>0.99p(1)$} & \textcolor{white}{$0.9-0.99p(1)$} & \textcolor{white}{$0.5-0.9p(1)$} & \textcolor{white}{$0.1-0.5p(1)$} & \textcolor{white}{$0.01-0.1p(1)$} & \textcolor{white}{$<0.01p(1)$}\\
			\midrule[0pt]

			200  & 12  & 4  & 5  & 6   & 4  & 169\\
			\midrule
			
			400  & 28  & 5  & 8  & 8   & 7  & 144\\
			
			\midrule
			
			600  & 45  & 7  & 9  & 9   & 9  & 121\\
			
			\midrule
			
			800  & 62  & 8  & 11  & 11   & 9  & 99\\
			
			\midrule
			
			1,000  & 80  & 9  & 12  & 12   & 11  & 76\\
			
			\midrule
			
			1,200  & 97  & 11  & 13  & 13   & 12  & 54\\
			
			\midrule
			
			1,400  & 115  & 12  & 14  & 14   & 13  & 32\\
			\midrule
			
			1,600  & 134  & 12  & 15  & 15   & 13  & 11\\
			
			\midrule
			
			1,800  & 152  & 13  & 16  & 16   & 3  & 0\\
			
			\midrule
			
			2,000  & 170  & 14  & 16  & 0   & 0  & 0\\
			\midrule
			
			2,200  & 189  & 11  & 0  & 0   & 0  & 0\\
			
			\midrule
			
			2,300  & 198  & 2  & 0  & 0   & 0  & 0\\
			
				\midrule
			
			2,400  & 200  & 0  & 0  & 0   & 0  & 0\\
			
		\end{dataTable2}
	\end{center}
	\caption{Results for $n=200$ and $r=0.1$}
	\label{table11}
\end{table}
}

We can think of the above as mimicking a situation where $R(1)$ is somewhere in Africa. Migration goes north to the middle East, and then either east into Asia or west into Europe. In each generation $90\%$ of the population stays and $10\%$ migrate a distance of $50-80$km. The migration span between $R(1)$ and $R(200)$ will then be between 10,000km and 16,000km. Table 1 shows, that, both in the bottleneck Scenario 1 and in Scenario 2, where the population size is constant over time in each region - after 2,400 generations, almost all genes in all regions have originated in region $R(1)$ in Africa. From these considerations and the considerations of Sections 2 and 3, where we connect the stability of the living conditions in a region to its reproductivity, we get the following:

\textit{Given that a region in Africa has over the last 50,000-100,000 years had more stable living conditions than other regions in the world, we can expect the majority of genes in today's human population to have originated in that region. This does not depend on whether or not there has been a population bottleneck. That also explains, why the genetic variation in today's human population is surprisingly small, i.e., observation (1.) above. The explanation of observation (2.) is analogous.}

Next, in Table 2, we will consider a shorter time perspective with 44 generations (approximately 1,000 years). We will consider 3 cases of reproductive advantage, $r=0.3$, $r=0.4$ and $r=0.5$. And we will consider the 3 cases of 10 regions, 20 regions and 40 regions. We will just compute how much of the genes from $R(1)$ (region 1) have spread through the other regions after 44 generations. 

Table 2 shows that, with 10 regions, already with a reproductive surplus of $r=0.3$ (on average $2.6$ children per individual) for $R(1)$, there is, over a timespan of 1,000 years, little difference between the bottleneck case and the case, when all regions have the same population size all the time. With 20 or 40 regions the difference is bigger, even with $r=0.4$ or $r=0.5$.
{\small
\begin{table}[h!]
	\begin{center}
		\begin{dataTable2}%
			{@{\hspace{2ex}} c @{\hspace{2ex}} c @{\hspace{2ex}} c @{\hspace{2ex}} c @{\hspace{2ex}} c @{\hspace{2ex}} c@{\hspace{2ex}} c@{\hspace{2ex}} c @{\hspace{2ex}}}%
			
			\multirow{2}{2.1cm}{\textcolor{white}{Total number\\ of Regions}} & \multicolumn{6}{c}{\textcolor{white}{Number of regions where, after $44$ generations, the amount of genes originationg in $R(0)$ is:}}\\
			\cline{2-7}
			& \textcolor{white}{$>0.99p(1)$} & \textcolor{white}{$0.9-0.99p(1)$} & \textcolor{white}{$0.5-0.9p(1)$} & \textcolor{white}{$0.1-0.5p(1)$} & \textcolor{white}{$0.01-0.1p(1)$} & \textcolor{white}{$<0.01p(1)$}\\
			\midrule[0pt]
			
			\midrule[1pt]
			
			 &   &   & $r=0.3$  &    &   & \\
			\midrule[0.5pt]
			
			10 & 8  & 2  & 0  & 0   & 0  & 0\\
			\midrule[0.5pt]
			
			20  & 8  & 2  & 4  & 4   & 2  & 0\\
			
			\midrule[0.5pt]
			
			40  & 8  & 2  & 4  & 4   & 3  & 19\\

			\midrule[1pt]
		
			 &   &   & $r=0.4$  &    &   & \\
			 
			\midrule[0.2pt]
			
			10  & 10  & 0  & 0  & 0   & 0  & 0\\
			
			\midrule[0.5pt]
			
			20  & 11  & 3  & 5  & 1   & 0  & 0\\
			
			\midrule[0.5pt]
			
			40  & 11  & 3  & 5  & 4   & 3  & 14\\
			
			\midrule[1pt]
			 &   &   & $r=0.5$  &    &   & \\
			\midrule[0.5pt]
			
			10  & 10  & 0  & 0  & 0   & 0  & 0\\
			\midrule[0.5pt]
			
			20  & 15  & 4  & 1  & 0   & 0  & 0\\
			
			\midrule[0.5pt]
			
			40  & 15  & 4  & 4  & 4   & 4  & 9\\

		\end{dataTable2}
	\end{center}
	\caption{}
	\label{table22}
\end{table}
}

We will now make a numerical study of a type of Scenario 2, to estimate how much time it takes for genes originating in $R(1)$ to dominate a region, once they have started to enter a region. So, we assume $p(1)=p(2)= \ldots =p(n)$. We define the transition period of $R(m)$ to be the time period, when at least $0.01 p(1)$ but not more than $0.99 p(1)$ of the genes in $R(m)$ originate in $R(1)$. In Table 3, $g$ is the number of generations it takes for genes originating in $R(1)$ to make up $0.01 p(1)$ of the genes in $R(m)$ and $h$ is the number of generations it takes to make up $0.99 p(1)$  of the genes in $R(m)$. $h - g$ is the length of the transition period in generations. In Table 3 we have multiplied by 23 and rounded off to give the length of the transition period in years. Also we put the reproductive surplus $r$ to be $0.1$ and the number of regions $n$ to be $100$.

\begin{table}[h!]
	\begin{center}
		\begin{dataTable}%
			{@{\hspace{2ex}} l @{\hspace{6ex}} r @{\hspace{6ex}} r @{\hspace{6ex}} r @{\hspace{2ex}}}%
			{\textcolor{white}{Region}}&
			\textcolor{white}{$g$}&
			\textcolor{white}{$h$}&
			{\textcolor{white}{Length of transition period in years}}\\ \midrule[0pt]
			$R(1)$ & 0 & 0 & 0 \\ 
			\midrule
			$R(10)$ & 38 & 170 & 3,000 \\ 
			\midrule
			$R(20)$ & 108 & 300 & 4,400 \\ 
			\midrule
			$R(30)$ & 185 & 423 & 5,500 \\ 
			\midrule
			$R(40)$ & 266 & 542 & 6,300 \\ 
			\midrule
			$R(50)$ & 350 & 658 & 7,100 \\ 
			\midrule
			$R(60)$ & 435 & 773 & 7,800 \\ 
			\midrule
			$R(70)$ & 521 & 887 & 8,400 \\ 
			\midrule
			$R(80)$ & 608 & 1,000 & 9,000 \\ 
			\midrule
			$R(90)$ & 696 & 1,112 & 9,600 \\ 
			\midrule
			$R(100)$ & 784 & 1,223 & 10,100 \\ 	
		\end{dataTable}
	\end{center}
	\caption{}
	\label{table13}
\end{table}
We can think of Table 3 as mimicking a situation where R(1) is populated by early moderns and $R(2)$, $R(3)$, $\ldots, R(100)$  are originally populated by European Neandertals. Migration in Europe goes from east to west. Then we can think of the transition period as the period of coexistence between moderns and Neandertals. Table 3 suggests that the length of this period increases as we move further to the west in Europe. This is consistent with the results in \cite{Trinkausetal} and \cite{Trinkaus}.

\begin{remark}\label{rem32}
{\rm If we change the matrix $M(r)$ above to $(1+d) M(r)$, we get a population increase by the factor $(1+d)$ in each generation. This is uniform over all regions. And it does not affect Tables 1, 2 and 3.}
\end{remark}

\section{Some concluding remarks}
There are many ways to extend the study in sections 1 through 4 above. One is to study more complicated migration patterns than just migration $r$ from $R(m)$ to $R(m+1)$ (one-sided migration). As a natural generalization, one could also consider some migration $q$ from $R(m)$ to $R(m-1)$ (two-sided migration). Some numerical experiments indicate that the one-sided migration with $r=0.1$ has big similarities with two-sided migration where $r-q=0.1$. In the last case migration goes ``on average'' from $R(m)$ to $R(m + 1)$.
In the model described in this paper we  have considered that most of the regions are reproductively neutral. Hence, in another generalization we might consider many regions that are not reproductively neutral. Some numerical experiments indicate that spreading out a reproductive deficit over several regions will somewhat increase the transition times, but not essentially change the big picture.
A deeper understanding of the genetic mechanisms behind lower (male) fertility among hybrids of closely related species would, of course, be important, in order to include such considerations in a study of population genetics.


\bigskip 

\noindent \paragraph{\textbf{Acknowledgements.}} The second and third authors were supported by Grant PGC2018-097286-B-I00.



\begin{bibdiv}
	\begin{biblist}
	
\bib{Enflo}{article}{
	author={P. H. Enflo},
	author={J. Hawks},
	author={M. Wolpoff},
	title={A simple reason why Neandertal ancestry can be consistent with current DNA information},
	journal={American Journal of Physical Anthropology},
	volume={114},
	year={2001},
	pages={Suplement 32:62},
}

	\bib{NEW-11}{article}{
	author={L. Chikhi},	
	author={G. Destro-Bisol},
	author={G. Bertorelle},
	author={V. Pascali},
	author={G. Barbujani},
	title={Clines of nuclear DNA markers suggest a largely Neolithic ancestry of the European gene pool},
	journal={PNAS},
	volume={95},
	year={1998, July},             
	pages={9053 - 9058},
} 

\bib{Trinkausetal}{article}{
	author={C. Duarte},
	author={J. Maur\'icio},
	author={P. Pettit},
	author={P. Souto},
	author={E. Trinkaus},
	author={H. Plicht},
	author={J. Zilh\~{a}o},
	title={The early Upper Paleolithic human skeleton from the Abrigo do Lagar Velho (Portugal) and modern human emergence in Iberia},
	journal={Proc. Natl. Acad. Sci. USA},
	volume={96},
	number={13},
	year={1999},
	pages={7604 - 7609},
}

\bib{Frayer}{collection}{
	author={D. Frayer},
	title={The persistence of Neandertal features in post-Neandertal Europeans in: Continuity or Replacement: Controversies in Homo sapiens Evolution},
	editor={G. Br\"{a}uer},
	editor={F. H. Smith},
	date={1992},
	part={pp. 179\ndash 188},
}

\bib{Hartl}{book}{
	author={D. Hartl},
	author={A. Clark},
	title={Principles of Population Genetics},
	publisher={Sinauer Associates, Inc},
	edition={Third Edition},
	date={2006},
}

\bib{NEW-5}{article}{
	author={M. Krings},
	author={A. Stone},
	author={R.W. Schmitz},
	author={H. Krainitzki},
	author={M. Stoneking},
	author={S. P\"{a}\"{a}bo},
	title={Neandertal DNA sequences and the origin of modern humans},
	journal={Cell},
	date={1997, Jul 11},
	volume={90 (1)},
	pages={19 - 30},
}
{\color{black}
\bib{nueva}{article}{
author={M. Kuhlwilm},
author={et al.},
title={Ancient gene flow from early moderns into Eastern Neanderthals},
journal={Nature},
date={2016},   
doi={10.1038/nature 16544},    
}
}

\bib{Morin}{article}{
	author={E. Morin},
	title={Evidence for declines in human population densities during the early Upper Paleolithic in western Europe},
	journal={PNAS},
	volume={105},
	year={2008},
	pages={48 - 53},
}


	\bib{Paabo2014}{article}{
	author={S. Sankararaman},
	author={S. Mallick},
	author={M. Dannemann},
	author={K. Pr\"ufer},
	author={S. P\"{a}\"{a}bo},
	author={N. Patterson},
	author={D. Reich},
	title={The landscape of Neandertal ancestry in present-day	humans},
	journal={Nature},
	date={2014},    
	volume={507 (7492)},
	pages={354 - 357},         
}  

\bib{Sankararaman2016}{article}{
	author={S. Sankararaman},
	author={S. Mallick},
	author={N. Patterson},
	author={D. Reich},
	title={The Combined Landscape of Denisovan and Neanderthal Ancestry in Present-Day Humans},
	journal={Current Biology},
	date={2016},    
	volume={26 (9)},
	pages={1241 - 1247},     
	DOI={http://dx.doi.org/10.1016/j.cub.2016.03.037},    
}  

	\bib{NEW-10}{article}{
	author={J. D. Wall},	
	author={M. A. Yang},
	author={F. Jay},
	author={S. K. Kim},
	author={E. Y. Durand},
	author={L. S. Stevison},
	author={C. Gignoux},
	author={A.Woerner},
	author={M. F. Hammer},
	author={M. Slatkin},
	title={Higher Levels of Neanderthal Ancestry in East Asians than in Europeans},
	journal={Genetics},
	volume={194},
	year={2013, May},             
	pages={199 - 209},
} 

\bib{Trinkaus}{article}{
	author={J. Zilh\~ao},
	author={E. Trinkaus},
	title={Portrait of the artist as a child. The Gravettian human skeleton from the Abrigo do Lagar Velho and its archaeological context},
	journal={Trabalhos de Arqueologia},
	volume={22},
	year={2002},
	pages={pp. 610},
}


		
	\end{biblist}
\end{bibdiv}
\end{document}